\documentclass[apjl]{emulateapj}
\usepackage{apjfonts}
\usepackage{graphicx}
\usepackage{amsmath}
\usepackage{amssymb}
\usepackage{amsfonts}
\usepackage{lmodern}
\shorttitle{ULXs at sub-Eddington Luminosities}
\shortauthors{Burke et al.}

\begin{document}

\newcommand{\lerg}[1]{10^{#1}~{\rm erg~s^{-1}} }
\newcommand{\msol}{{\rm ~M_\odot}}
\newcommand{\mdot}{\dot{M}}

\title{The fading of two transient ULXs to below the stellar mass Eddington limit}

\author{Mark~J.~Burke\altaffilmark{1,2},
Ralph~P.~Kraft\altaffilmark{2}, 
Roberto~Soria\altaffilmark{4},
Thomas~J.~Maccarone\altaffilmark{6},
Somak~Raychaudhury\altaffilmark{1,3},
Gregory~R.~Sivakoff\altaffilmark{5},
Mark~Birkinshaw\altaffilmark{7},
Nicola~J.~Brassington\altaffilmark{8},
William~R.~Forman\altaffilmark{2},
Martin~J.~Hardcastle\altaffilmark{8},
Christine~Jones\altaffilmark{2},
Stephen~S.~Murray\altaffilmark{9,2} and
Diana~M.~Worrall\altaffilmark{7}
}

\altaffiltext{1}{School of Physics and Astronomy, University of
  Birmingham, Edgbaston, Birmingham, B15 2TT, UK 
\email{mburke@star.sr.bham.ac.uk}}

\altaffiltext{2}{
  Harvard-Smithsonian Center for Astrophysics, 60 Garden Street,
  Cambridge, MA 02138, USA}

\altaffiltext{3}{Department of Physics, Presidency University, Kolkata 700 073, India}

\altaffiltext{4}{International Centre for Radio Astronomy Research, Curtin University, GPO Box U1987, Perth, WA 6845, Australia}

\altaffiltext{5}{Department of Physics, University of Alberta,
Edmonton, Alberta T6G 2E1, Canada}

\altaffiltext{6}{Astronomy \& Astrophysics Group, Texas Tech University, Lubbock,TX 79409-105, USA}

\altaffiltext{7}{HH Wills Physics Laboratory, University of Bristol, Tyndall Avenue, Bristol BS8 1TL, UK}

\altaffiltext{8}{School of Physics, Astronomy, and Mathematics, 
			University of Hertfordshire,
			Hatfield, AL10 9AB, UK}

\altaffiltext{9}{Department of Physics and Astronomy, Johns Hopkins University, 3400 N. Charles Street, Baltimore, MD 21218, USA}

\begin{abstract}
We report new detections of the two transient ultraluminous X-ray sources (ULXs) in NGC 5128 from an ongoing series of \emph{Chandra} observations. Both sources have previously been observed $L_x(2-3) \times \sim \lerg{39}$, at the lower end of the ULX luminosity range. The new observations allow us to study these sources in the luminosity regime frequented by the Galactic black hole X-ray binaries (BH XBs).

We present the recent lightcurves of both ULXs. 1RXH J132519.8-430312 (ULX1) was observed at $L_x \approx 1\times\lerg{38}$, while CXOU J132518.2-430304 (ULX2) declined to $L_x \approx 2\times\lerg{37}$ and then lingered at this luminosity for hundreds of days.  We show that a reasonable upper limit for both duty cycles is 0.2, with a lower limit of 0.12 for ULX2.  This duty cycle is larger than anticipated for transient ULXs in old stellar populations. 

By fitting simple spectral models in an observation with $\sim50$ counts we recover properties consistent with Galactic BH XBs, but inconclusive as to the spectral state.  We utilise quantile analyses to demonstrate that the spectra are generally soft, and that in one observation the spectrum of ULX2 is inconsistent with a canonical hard state at $>95\%$ confidence.  This is contrary to what would be expected of an accreting IMBH primary, which we would expect to be in the hard state at these luminosities.  

We discuss the paucity of transient ULXs discovered in early-type galaxies and excogitate explanations.  We suggest that the number of transient ULXs scales with the giant and sub-giant populations, rather than the total number of XBs.

\end{abstract}


\section{Introduction}
\label{s:intro}

Ultraluminous X-ray sources (ULXs) are X-ray point sources with isotropic luminosities $>\lerg{39}$, displaced from the nucleus of their parent galaxy.  Our knowledge of such systems has increased dramatically in the \emph{Chandra} and \emph{XMM-Newton} eras, where high spatial resolution coupled with large variability has supplied strong evidence against the high luminosities resulting from source superposition.  These luminosities are an order of magnitude above the Eddington limit $L_{Edd}$ of accreting neutron stars (NSs) and therefore it seems likely that ULXs possess black hole (BH) primaries.  

The key debate in ULX research concerns the mass of the compact object.  Stellar mass black holes, which have an empirical mass distribution peaked at $\sim 8 {\rm M_{\odot}}$ \citep{2010ApJ...725.1918O,2012ApJ...757...36K}, would have to be emitting $\gtrsim L_{Edd}$, while some analyses favour an explanation involving intermediate mass black holes (IMBHs) of $M \gtrsim 10^3 {\rm M_{\odot}}$ accreting at a fraction of their Eddington rate.  The existence of IMBHs has implications across a broad range of astrophysics including the studies of stellar populations, gravitational wave astronomy and the origins of super massive black holes.

\subsection{Transient ULXs}
\label{sec:transulx}

Most ULXs are observed to be persistent over multi-epoch observations, and consistently at luminosities above $\lerg{39}$  \citep{2004AAS...204.4904C,2006ApJ...642..171L}.  However, we are now aware of a handful of transient ULXs that experience large long-term variability over many orders of magnitude, rising to $L_x>\lerg{39}$ from below the typical detection limits of extragalactic X-ray binaries (XBs).   The study of such transient sources is important because they must cross the luminosity range typical of Galactic BH XBs during their rise and decay.  If a ULX can be observed during such a transition we can attempt to determine whether it behaves like an ordinary stellar-mass BH, in terms of its spectral transitions and timing properties,  or whether it is a fundamentally different type of accretion phenomenon in the ordinary BH luminosity regime.  

Few transient ULXs are known due to the lack of regular deep X-ray monitoring for every galaxy over a prolonged period.  It may prove to be the case that outburst durations are $\sim10-100$ years, particularly for sources associated with old stellar populations \citep{2002ApJ...571L.103P}, which is longer than the \emph{Chandra} and \emph{XMM} eras.  A handful of sources have now shown variability and outburst timescales comparable to Galactic BH LMXBs, with luminosities on occasion $\gtrsim \lerg{39}$ and $\lesssim \lerg{38}$ at other times.  These ULXs have been detected in both starburst \citep[e.g.][]{0004-637X-668-2-941,2012ApJ...750..152S,2005A&A...442..925B} and early-type \citep[e.g.][]{2001ApJ...560..675K,2012ApJ...760..135R} galaxies.     Outburst durations and recurrence times remain largely unconstrained, but some outbursts have remained bright ($>\lerg{39}$) for over a year \citep{2012ApJ...750..152S}.  Seemingly persistent behaviour is a common feature of bright ($>8\times\lerg{38}$) sources in early-type galaxies   \citep[e.g.][]{2006MNRAS.371.1903I}, and outside of globular clusters the high accretion rates implied by such luminosities suggests that the sources may have started to accrete comparatively recently. An empirical, scattered correlation between outburst duration and orbital period has been demonstrated for Galactic XBs \citep{2004MNRAS.355..413P} and a stronger correlation found between the peak $L_x$ and orbital period by \citep{2010ApJ...718..620W}.  This leads us to predict that the population of bright low mass X-ray binaries is dominated by large period ($>30$d) systems with an evolved companion, experiencing large outbursts with larger recurrence times.

The physical origin of the distinction between the transient ULXs and the persistent sources remains unknown.  The majority of Galactic BH low mass X-ray binaries (LMXBs) are transient, experiencing outbursts on timescales of weeks to months, recurring on months to years \citep[for review see~][]{2006ARA&A..44...49R}.    It is thought that transient behaviour is the result of the disk instability mechanism (DIM) \citep{2001NewAR..45..449L}  which has been successful in explaining the limit-cycle behaviour observed from the lightcurves of low-period cataclysmic variables \citep{1984PASP...96....5S}.  Transient behaviour should occur from below a critical accretion rate, above which the entire disk is always warmer than the  ionisation temperature of hydrogen ($\approx6,500$~K) and emission is persistent. In LMXB, irradiation of the outer disk by the central X-ray source should mean that this threshold is at a lower accretion rate than for CVs \citep{2001A&A...373..251D} and less than a few percent Eddington,~a prediction that was recently found to be in good agreement with the accretion of Galactic sources \citep{2012MNRAS.424.1991C}. A question that remains regarding transient ULXs is whether they are IMBHs that outburst via the DIM and observed at a fraction of their $\dot{M}_{Edd}$, or ordinary stellar BHs accreting $>\dot{M}_{Edd}$.   \cite{2003A&A...409..697M} showed that the transition from the soft to the hard state before quiescence occurs at a few percent Eddington for Galactic BH XBs, and \cite{2003MNRAS.345L..19M} showed that this is also the case for super massive BH accretion in AGNs, and also that the low/hard and quiescence are indistinguishable \citep[albeit softening slightly with decreasing $L_x$,][]{2009ApJ...705.1336C}.  This similar behaviour observed at the extremes of the BH mass spectrum suggest a common accretion behaviour that should be applicable to the complete range of BH masses. An IMBH of $1000 {\rm M_{\odot}}$ would have $L_{Edd}\sim 1.5 \times \lerg{41}$, and therefore for $L_x \lesssim \lerg{39}$ the accretion rate is much less than 1\% Eddington, and the source spectra should be hard.   In contrast, BH XBs at peak $L_x$ are usually in the so called steep power law (aka very-high) state, before dimming into the canonical thermal dominant state, when the thermal emission from the disk dominates the spectrum.  Therefore evidence of a soft or disc-dominated spectrum at sub-UL luminosities argues against accreting IMBHs. It is for this reason that Hyper-luminous X-ray source HLX-1 has been dubbed the `best' IMBH candidate to date as the spectra are consistent with canonical thermal dominant and hard states at $\lerg{42}$ and $\lerg{40}$, respectively \citep{2011ApJ...743....6S}.

The identification and classification of ULX optical counterparts are problematic, as observations of the host galaxies are often unable to resolve the dense stellar fields, nor deep enough to place meaningful constraints on the color via a significant non-detection.  The greater abundance of ULXs in late-type galaxies, particularly star-forming regions \citep{2009ApJ...703..159S}, suggests young donors may reside in the majority of systems.   Those companions that are identified are often blue, which may indicate an OB donor \citep[e.g.][]{2004ApJ...602..249L}. However, for some sources it has been shown that the blue light is only present during the X-ray outburst, which suggests that the optical light is mostly reprocessed emission in the disk \citep[e.g.][]{2012ApJ...750..152S}.  Recently, an \emph{HST} census of optical counterparts in nearby ($\le 5 $ Mpc) galaxies \citep{2013arXiv1303.1213G} reported upper limits of $M_v=-4$ to $-9$ for 9/33 systems, ruling out O type companions for 4, and used SED fitting to rule out a further 20 O-types and an OB donor completely for one ULX in NGC 253.    The nature of the counterpart is important in the context of the IMBH debate, as massive donors should produce persistent X-ray sources unless the compact object is very massive \citep{2001ApJ...552L.109K}, which is the dichotomy between BH HMXB and BH LMXB donors observed in the Milky Way \citep{2006ARA&A..44...49R}. For transient behaviour to occur for $M\approx5-20\msol$ donors from a young stellar population, the compact object must be $>50\msol$ \citep{2004ApJ...603L..41K}.  A ULX with a B9 supergiant donor in spiral galaxy NGC 7793 that was recently observed to vary between $L_x\sim(5-400)\times\lerg{37}$ (Motch et al., in prep.) would have violated this hypothesis, as the BH mass is constrained $\le 15 \msol$. However, the authors argue that this is not a true transient, but rather that it is a high inclination source where the fainter state results from obscuration by the precessing disk rim.

\subsection{NGC 5128 ULXs}
NGC 5128 (Centaurus A) is the nearest optically luminous early-type galaxy, situated at a distance of 3.7 Mpc \citep{2007ApJ...654..186F}, with $M_B = -21.1$.  Two ULXs are known in Cen A, both of which are transients.  The first ULX (1RXH J132519.8-430312, herein ULX1) was discovered by \emph{ROSAT} \citep{2000A&A...357L..57S} and re-detected near the start of the \emph{Chandra} era \citep{2001ApJ...560..675K,2006ApJ...640..459G}.  The Chandra detections showed a soft spectrum, well described by a cool multi-temperature disk blackbody or a steep power law, and from this we infer $L_x \gtrsim \lerg{39}$.  The luminosity may have been as high as $5\times\lerg{39}$ in the ROSAT HRI observations.  

Six 100~ks \emph{ Chandra} observations of NGC\,5128 were taken in 2007 as
part of the Cen A Very Large Project (VLP), spanning the course of 2
months \citep{2007ApJ...671L.117J}.  Over the course of these observations
a second ULX was discovered \citep[CXOU J132518.2-430304, herein
ULX2,][]{2008ApJ...677L..27S}. Its spectra were reminiscent of the steep power
law state; it softened during the course of the outburst, as would be
expected for a BH LMXB entering the thermal dominant state.  The spectra were modelled using thermal disk blackbody and power law
components.  Returning to study these spectra, we note that the
inner-temperature of the disk is cool ($0.6-1.0$ keV) for a stellar mass BH at such a high
luminosity, and the normalisation of the component requires an inner disk
radius of $\approx 19$ km for a face-on disk, much smaller than the
innermost stable circular orbit for a $\sim10 \msol$ BH.  Therefore the
disk component is unlikely to be a physical description of the actual
disk, and the source has spectra consistent with either the steep-power law or the so-called `Ultraluminous state' \citep{2009MNRAS.397.1836G}.  Unambiguously distinguishing the ultraluminous state relies on high quality data with a enough sensitivity above 5 keV to detect the high-energy rollover with simultaneous low temperature disk blackbody in the spectrum.  This feature is hard to detect with \emph{Chandra}, which has a smaller collecting area than \emph{XMM-Newton}.

Previously, we have shown that the transient BH LMXB candidates of Cen A, analogous to the BH systems of the Milky Way in terms of transient behaviour and thermal state spectra, are only associated with a merged late-type remnant \citep{2013ApJ...766...88B}.  We suggested that the merged late-type was a more favourable environment for such systems to exist owing to the ancient stellar population of the halo \citep[$\sim 12$ Gyrs, ][]{2011A&A...526A.123R} resulting in a relative paucity of massive enough donors for the longer orbital periods required to achieve the observed luminosities.  ULX1 and ULX2 are in the halo of Cen A, to the south-west of the nucleus, and are therefore associated with an older stellar population than the more typical (in Milky Way terms) BHC LMXBs that are associated with the late-type galaxy remnant.      

In this paper we report on a series of Chandra observations since the 2007 VLP.  We use the VLP observations to show the lowest upper limit for the quiescent state of ULX1, and report on subsequent detections of both ULXs.  Both sources are detected at luminosities substantially sub-Eddington for a 10 ${\rm M_\odot}$ BH LMXB, and we take advantage of a rare opportunity to study the spectra of ULXs during their decline from outburst.  We discuss spectra, recurrence times and duty cycles and show that these are further evidence of the connection between ordinary stellar mass BH LMXBs and ULXs, and argue against the need for IMBH primaries in these systems.  Finally, we discuss the lack of transient ULX discoveries in early-type galaxies.

\begin{deluxetable*}{lcccccccccccc}
\tabletypesize{\scriptsize}
\tablecaption{Net Counts (0.5-8.0 keV)}
\tablehead{
\colhead{ULX} &
\colhead{10723} &
\colhead{10724} &
\colhead{10725} &
\colhead{10726} &
\colhead{10722} &
\colhead{11846} &
\colhead{11847} &
\colhead{12155} &
\colhead{12156} &
\colhead{13303} &
\colhead{13304} &
\colhead{15294}
}

\tablecolumns{13}
\startdata
1 & $<5.9$ &
$<2.6$ &
$<2.6$ &
$<5.9$ &
$<4.3$ &
$<3.2$ &
$<3.2$ &
$<3.2$ &
$\mathbf{24.2^{+9.7}_{-7.6}}$ &
$<2.6$ &
$<2.6$ &
$<2.6$ \\
2 & $\mathbf{20.1^{+9.6}_{-8.1}}$ &
$\mathbf{9.9^{+6.6}_{-4.6}}$ &
$\mathbf{6.5^{+5.6}_{-3.6}}$ &
$<4.9$ &
$\mathbf{ 45.5^{+13.5}_{-11.5}}$ &
$<3.8$ &
$\mathbf{ 4.3^{+4.8}_{-2.8}}$ &
$<3.2$ &
$\mathbf{ 8.6^{+6.3}_{-4.3}}$ &
$<3.5$ &
$<3.2$ &
$<2.6$ \\

\enddata
\tablecomments{{\bf Net counts} and $90\%$ upper limits for each new observation of ULX1 and ULX2.} 
\label{tab:COUNTS}
\end{deluxetable*}

\section{Data Preparation and Analysis}
Since 2007, Chandra observations have been made as part of the HRC Guaranteed Observation Time program (PI: Murray), amounting to a series of regular 5 ks `snapshots' and a single 50 ks observation (ObsID 10722).  We previously presented data from some of these observations in table 1 of \cite{2012ApJ...749..112B}, and in addition to this we now include three further datasets, obsIDs 13303, 13304 and 15294 made on MJD 56031, MJD 56168 and MJD 56387.  These data were reduced and aligned to the other Cen A \emph{Chandra} ACIS observations in the manner detailed by \cite{2013ApJ...766...88B}.  We make use of the VLP data to obtain deep upper limits on the emission from ULX1 during 2007.

\subsection{Observed Count Rates}
\label{sec:count}
\begin{figure*}
\begin{center}
\includegraphics[width=0.35\hsize, angle=270]{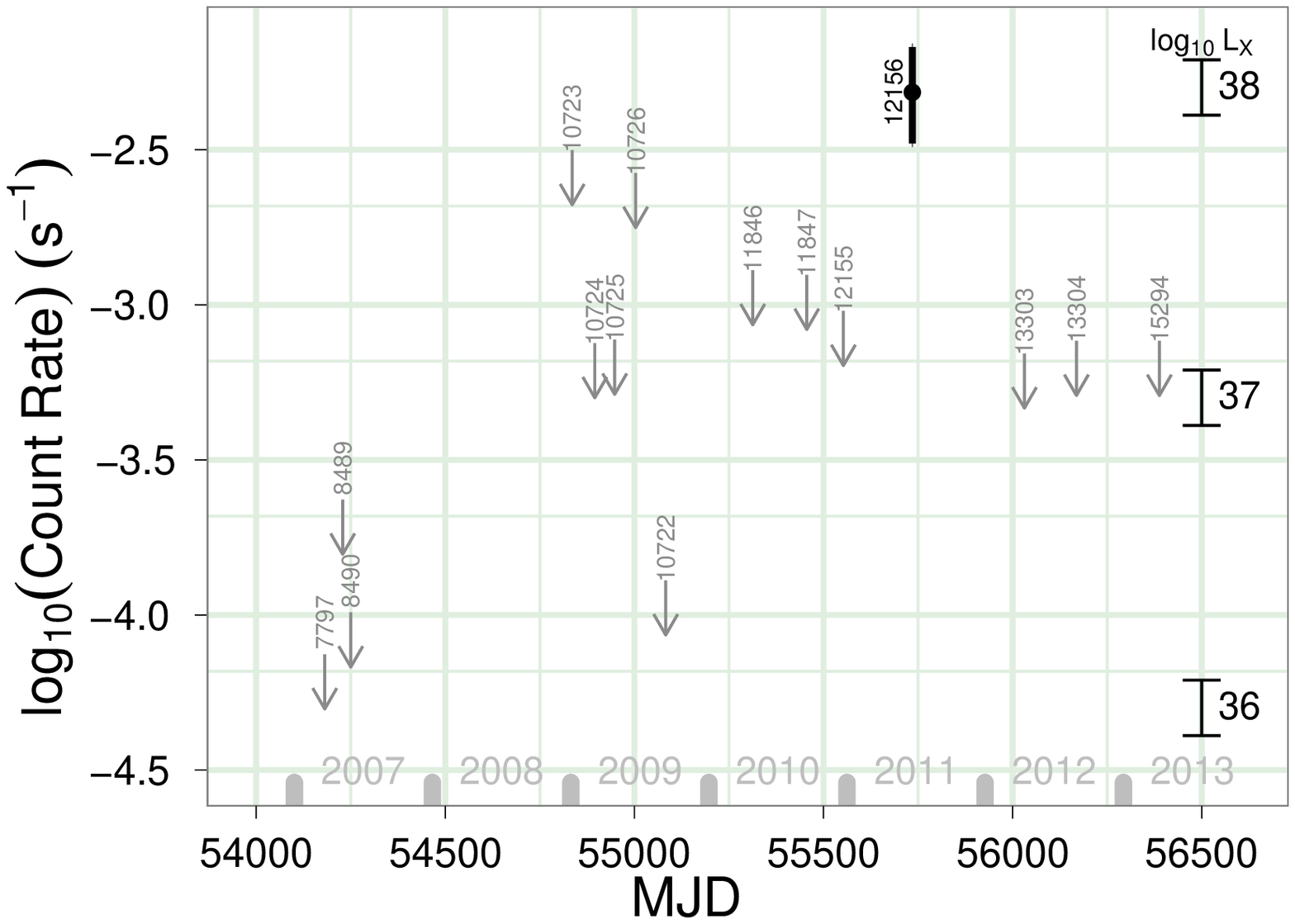} \includegraphics[width=0.35\hsize, angle=270]{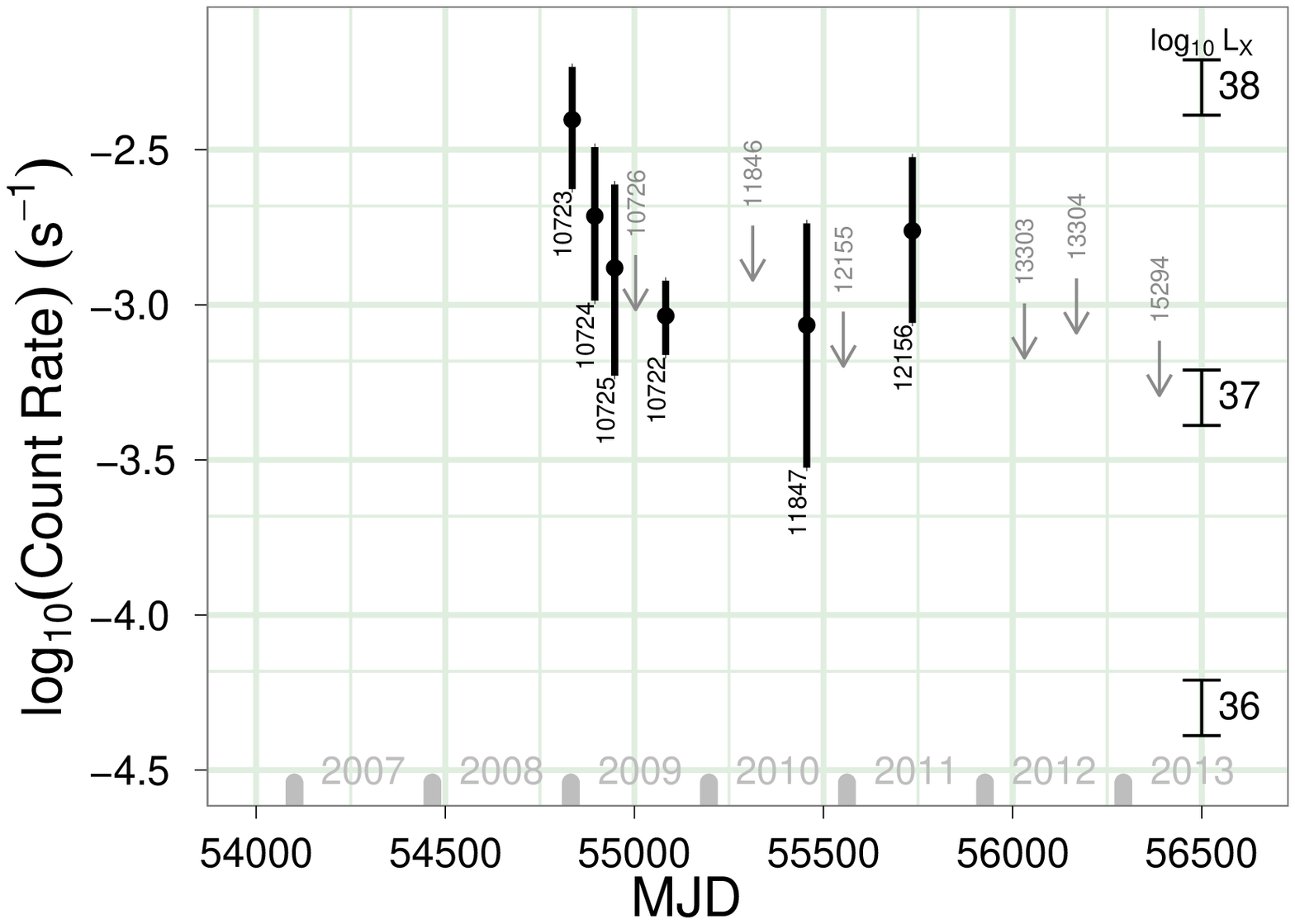}
\caption{X-ray lightcurves (0.5-8.0 keV) since 2007 for (\emph{left}) ULX1, including the VLP observations, and (\emph{right}) ULX2.  Arrows denote 90\% confidence upper limits and obsIDs are labelled.\label{fig:LC} We obtained a detection for ULX2 using obsIDs 11846, 11847 \& 12155 (Section~\ref{sec:count}).  On the right of each lightcurve we show ${\rm \log_{10}}(L_x)$, for a range of power law spectra with $\Gamma=1.5-2.5$.  Note that obsID 10722 is an ACIS-S observation and the $\log_{10}(L_x)$ were calculated using webPIMMs for ACIS-I observations.}
\end{center}
\end{figure*}

With the objective of securing count rates for the ULXs in the post-VLP observations, and upper limits for ULX1 in the 100 ks observations we used \emph{MARX}\footnote[1]{http://space.mit.edu/cxc/marx/} to define ellipses around the positions of both sources describing the 90\% PSF at 8 keV.  This was adequate to avoid source confusion in the snapshot observations, but in some of the 100 ks observations the positions are so much further off-axis that the position of ULX1 overlaps with a bright neighbour, so that source regions could only be defined for three of the VLP observations.  For background, we defined a source-free rectangular region based on the merged observation file, to the north of the ULXs.  We used the \emph{ciao} tool \emph{aprates} to estimate $90\%$ confidence count rates and upper limits, which we present as a lightcurve for both ULXs in  Fig.~\ref{fig:LC}.  The detection count rates are substantially smaller than when the sources were observed in the UL regime; $\approx (1-5) \times 10^{-3}~{\rm s^{-1}}$ down from $\approx8\times10^{-2}~{\rm s^{-1}}$ in the UL regime.  From the VLP observations, the lowest upper limit for ULX1 is $5\times10^{-5}~{\rm s^{-1}}$, which assuming a  power law described by photon index $\Gamma=1.7$ experiencing Galactic absorption, is an upper limit of $L_x \approx \lerg{36}$ over $0.5-8.0$ keV.

\begin{deluxetable}{lccccc}
\tabletypesize{\footnotesize}
\tablecaption{Hardness Ratios (0.5-1.5 keV)/(1.5-8.0 keV)}
\tablehead{
\colhead{ULX} &
\colhead{10723} &
\colhead{10724} &
\colhead{10725} &
\colhead{10722} &
\colhead{12156} 
}

\tablecolumns{6}
\startdata
1 & - & - & - & - & $1.20_{-0.45}^{+0.75}$ \\
2 & $1.62_{-0.82}^{+2.11}$ & $0.54_{-0.30}^{+0.66}$ & $0.55_{-0.38}^{+1.01}$ & $1.14_{-0.34}^{+0.50}$ & $0.63_{-0.38}^{+0.89}$ \\
\enddata
\tablecomments{Hardness ratios calculated using BEHRs \citep{2006ApJ...652..610P}. Confidence intervals are $68.5\%$} 
\label{tab:hard}
\end{deluxetable}

We combine three of the $\sim100$ ks observations (obsIDs 7797, 8489, 8490) to obtain the deepest upper limit to ULX1.  We obtain a count rate limit of $5.34\times 10^{-5}~{\rm s^{-1}}$, corresponding to a luminosity of $L_x\sim 1.2 \times\lerg{36}$.  Similarly we measure a combined upper limit of $1.6\times 10^{-4} {\rm s^{-1}}$ over obsIDs 13303, 13304 and 15294 subsequent to the latest detection, which means that $L_x<\lerg{37}$.

We combined obsIDs 11846, 11847 and 12155 to calculate average count rate for ULX2. Running \emph{wavdetect}, as before, identified the source in the combined image (there was no detection by wavdetect in obsID 11847 alone) and measure a count rate of $3.21_{-2.81}^{+3.93}\times 10^{-4}~{\rm s^{-1}}$ ($90\%$ confidence).  This would correspond to a ($0.5-8.0$ keV) luminosity of $5.8 \times \lerg{36}$, with an upper limit of $1.2\times \lerg{37}$.  Given the subsequent detection in obsID 12156, this suggests that the source remained in outburst after obsID 10722, lingering at $L_x\approx \lerg{37}$.  A combined measurement including obsID 12156 has a count rate of $6.6_{-3.2}^{+4.2} \times 10^{-4}~{\rm s^{-1}}$. We measure an average upper limit of $3.0\times10^{-4}~{\rm s^{-1}} $ ($L_x<\lerg{37}$) over obsIDs 13303, 13304 and 15294.

Finally, we tested the intra-observational variability of the sources, for each individual detection, using the \emph{ciao} tool \emph{glvary}\footnote[2]{http://cxc.harvard.edu/ciao/ahelp/glvary.html} \citep{1992ApJ...398..146G}.  No variability was detected, and the variability indices were $\le 1$.  Significant variability is defined as variability index of at least 6.  

\subsection{Spectral fitting}
\label{sec:spec}
\begin{figure}
\begin{center}
\includegraphics[height=1\hsize]{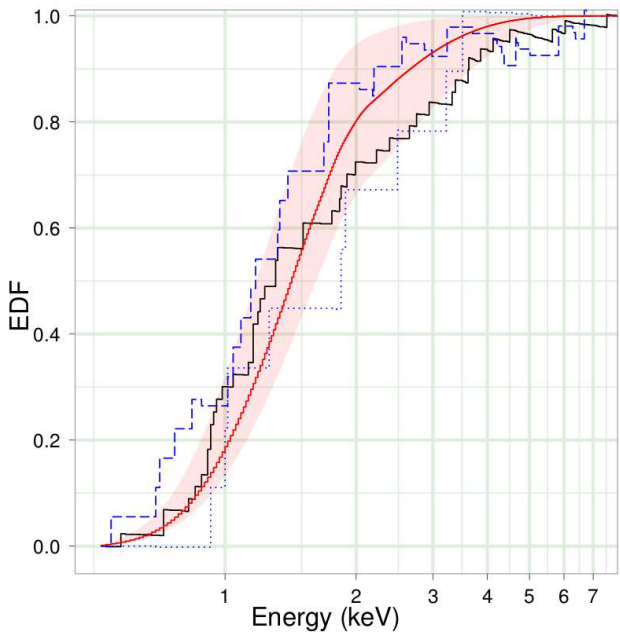}
\caption{Integrated counts (Empirical Distribution Function) from ULX2 in obsIDs 10722 (solid), 10723 (dashed) and 10724 (dotted), overlaid on the best fit $diskbb$ ($kT_{in}=0.74_{-0.33}^{+0.45}$ keV, see section~\ref{sec:spec}) from obsID 10722 (red line) with 90\% confidence region (shaded) \label{fig:QCCD}.}
\end{center}
\end{figure}

We attempted fitting simple spectral models to the more count-rich spectra.  ULX2 in obsID 10722 has $\sim45$ counts, and in obsID 10723 has $\sim22$ counts.  Using the modified Cash statistic \citep[][]{1979ApJ...228..939C,2011hxa..book.....A} (aka `W' statistic) in \small{XSPEC} we fit the ungrouped spectra using simple power law and disk blackbody models (\emph{powerlaw} $\&$ \emph{diskbb}, respectively).  In fitting the low count spectra in various joint fits we could not achieve a good fit, with  $95-100\%$ of Monte-Carlo goodness-of-fit simulations possessing a lower test statistic then that achieved in fitting.  For obsID 10722, we found converging fits with an absorbed power law model (\emph{phabs} $\times$ \emph{powerlaw}) and an absorbed disk-blackbody model (\emph{phabs}$\times$\emph{diskbb}).  We initially left the absorption parameter $N_H$ free, to make use of the diagnostics of \cite{2010ApJ...725.1805B} which we applied successfully in Burke et al. (2013) to infer the spectral states of many X-ray binaries in Cen A.  However, the $N_H$ recovered from fitting had large uncertainties that rendered such inference impossible, and the point estimates tended towards zero.  When $N_H$ was fixed to the Galactic value ($8.4 \times 10^{20} ~{\rm cm^{-2}}$) the $diskbb$ fit yields $kT_{in}=0.74_{-0.33}^{+0.45}$ keV with $L_x\approx(0.7-1.6)\times \lerg{37}$.  The fit is acceptable, as Kolmogorov-Smirnov \& Anderson-Darling tests\footnote[3]{Available in \small{XSPEC 12.8.0}} produce values that are consistent with simulations, i.e. `goodness test' of 2000 samples shows that $\approx40\%$ of realisations produce smaller test statistics.  We present the integrated counts of ULX2 during obsIDs 10722, 10723 and 10724 in figure~\ref{fig:QCCD}, together with the best fitting \emph{diskbb} model and confidence region for obsID 10722.  For an absorbed power law, $\Gamma=2.2_{-0.6}^{+0.7}$ and $L_x\approx (1.1-2.2)\times \lerg{37}$, with $\sim20\%$ realisations produced better values of the test statistic.

\subsection{Quantile Analysis}
We have a very rare opportunity to try to infer the spectral state of ULXs in the non-UL regime.  However, in our observations there are often too few counts to perform reliable spectral fitting (see section~\ref{sec:spec} for discussion).  Hardness ratios, the traditional workhorse of  X-ray astronomy, are dependent on the subjective choice of energy bands.  For reference, we present hardness ratios for these detections in table~\ref{tab:hard},  calculated using the BEHRs software\footnote[4]{http://hea-www.harvard.edu/AstroStat/BEHR/\#refs}.  The spectra with the most counts, obsIDs 10722, 10723 are the softest, and while the lower-quality spectra appear harder, they are not significantly so.  Quantile analysis maximises the use of all the photons detected.  Using the source and background regions described in section~\ref{sec:count}, we extracted the spectra from the observations where a given ULX was detected. We employ techniques and software described by \cite{2004ApJ...614..508H,2009ApJ...706..223H}, who showed that the median energy $E_m$ is a superior tool to hardness ratios for inferring spectral states.  Our 5 detections of ULX2 and new detection of ULX1 all have $< 50$ counts per observation.  We present the median energies and $1\sigma$ uncertainties from the individual detections in table~\ref{tab:quan}.

\cite{2009ApJ...706..223H} define what they call a `quantile color-color diagram' (QCCD) to separate sources into spectral groups.  The axes of these plots are an expression of $E_m$ (in terms of the bounds of the energy band, x-axis) and the ratio of the energy of the 25\% to the 75\% quantiles ($\times3$,~y-axis).  For our purposes this was not an informative diagram for comparing the spectra, as the uncertainties in the y coordinate were too large to determine if any spectral changes had occured.  We therefore focus on the behaviour of the median energy.

\begin{deluxetable}{lccccc}
\tabletypesize{\footnotesize}
\tablecaption{Median Energies (keV)}
\tablehead{
\colhead{ULX} &
\colhead{10723} &
\colhead{10724} &
\colhead{10725} &
\colhead{10722} &
\colhead{12156} 
}

\tablecolumns{6}
\startdata
1 & - & - & - & - & $1.44_{-0.21}^{+0.21}$ \\
2 & $1.17_{-0.17}^{+0.17}$ & $1.83_{-0.54}^{+0.54}$ & $1.39_{-0.28}^{+0.28}$ & $1.30_{-0.19}^{+0.19}$ & $1.56_{-0.23}^{+0.23}$ \\
\enddata
\tablecomments{Median energies of ULX1 and ULX2, with $1\sigma$ uncertainties as calculated by \cite{2004ApJ...614..508H}.} 
\label{tab:quan}
\end{deluxetable}

\begin{figure*}
\begin{center}
\includegraphics[width=0.42\hsize, angle=270]{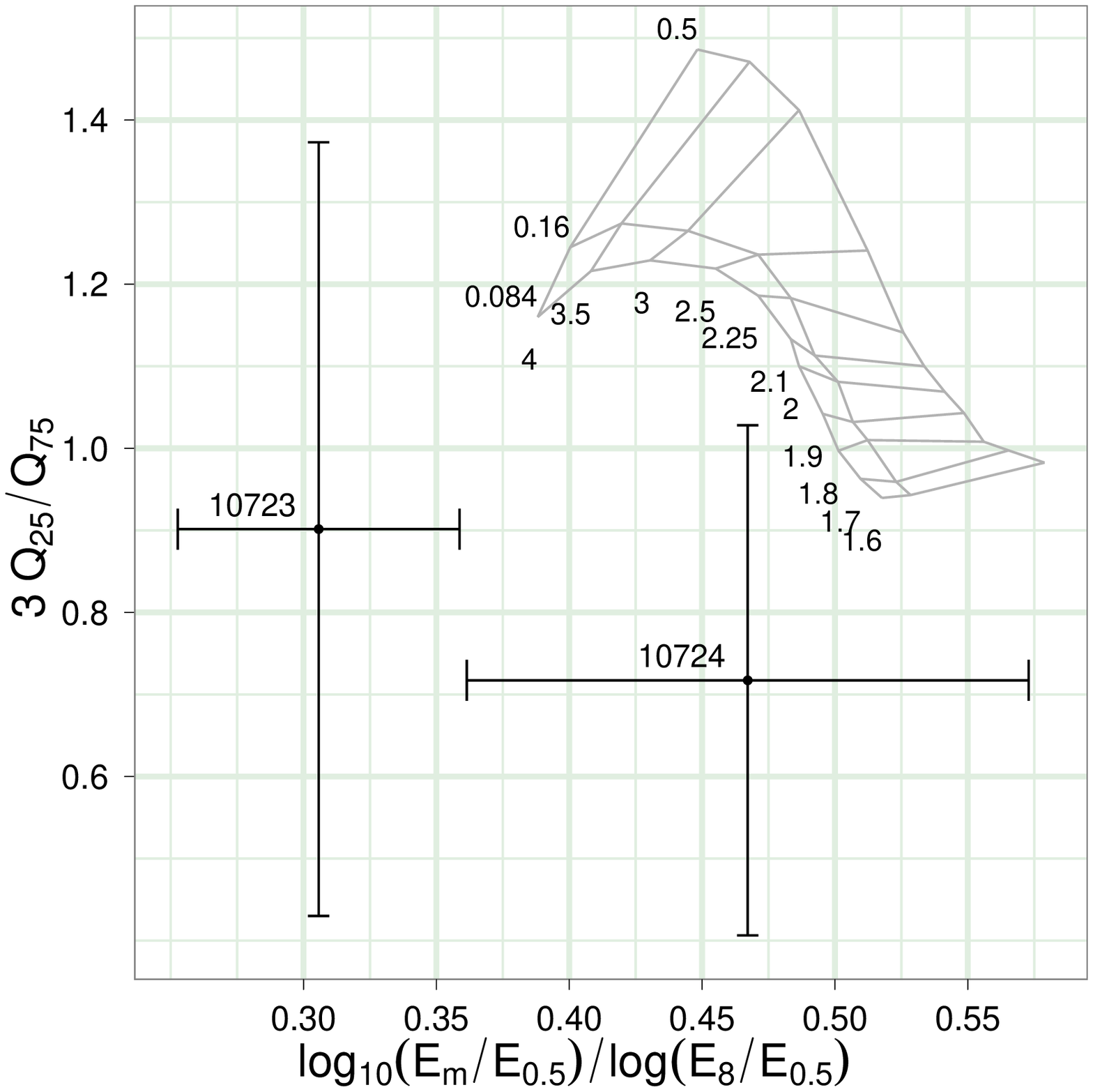} \includegraphics[width=0.42\hsize, angle=270]{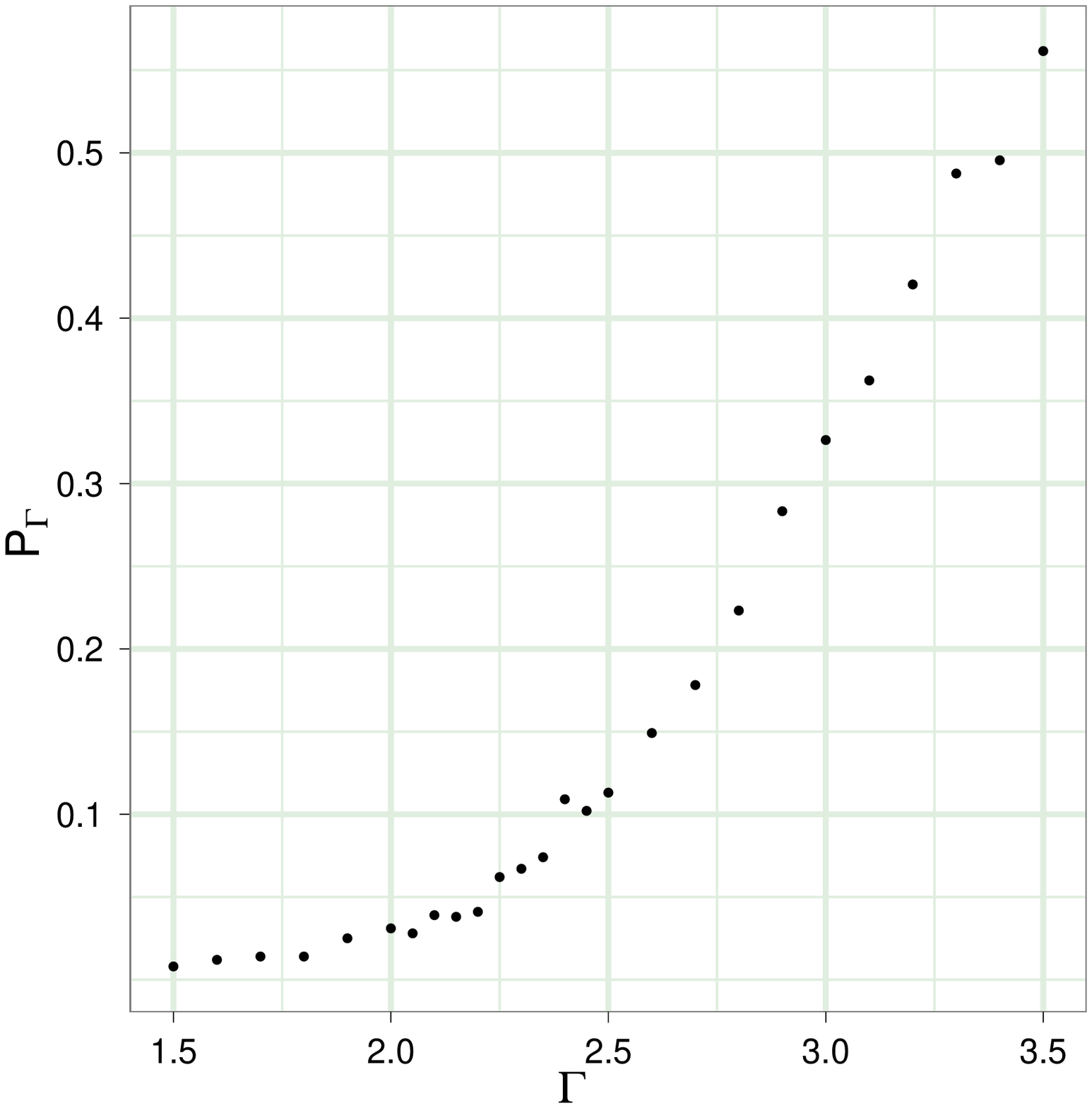}
\caption{\emph{Left}, QCCD for ULX2 with parameter grid corresponding to an absorbed power law for $N_H=(0.084, 0.16, 0.5)\times {\rm 10^{22}~cm^{-2}}$ and $\Gamma = 1.6-4.0$, with quantiles from obsIDs 10723 and 10724. \emph{Right}, proportion of simulated spectra that possess median energies as extreme as that measured for ULX2 in obsID 10723 for a range of $\Gamma$, with $N_H=0.084\times {\rm 10^{22}~cm^{-2}}$.  Each point represents a sample of 1000 simulated spectra. \label{fig:ULX2PG}}
\end{center}
\end{figure*}

A key issue that we wish to address is whether ULXs experience canonical BH LMXB spectral states in the classic luminosity regime.  As discussed in \S~\ref{sec:transulx}, opinions on this are divided into whether ULXs transition to a hard state (power law dominated) as they leave the UL regime (as would be expected for an IMBH, which would be at a few percent Eddington), where they remain in quiescence, or whether they enter a thermal dominant state first, as is seen for the majority of stellar mass BH systems.   To this end we include a small grid of parameters for absorbed power law spectra in Fig.~\ref{fig:ULX2PG}.  We present this next to the QCCD datum of ULX2 from obsID 10723, as this $E_m$ is at the most extreme position from the grid (see below).  \cite{2004ApJ...614..508H} state that the estimated uncertainties are possibly over-estimated by as much as $20-30\%$ in the $<30$ count regime, where most of our detections lie.  To determine how extreme this $E_m$ would be for a source in the hard state, we simulated many spectra described by a power law of $\Gamma=1.5$--$2.1$ with the same number of counts ($\approx22$) as obsID 10723, calculated $E_m$ for each spectrum, and then found the proportion $P_\Gamma$ of simulated spectra for each $\Gamma$ that had $E_m$ smaller than that measured from obsID 10723 (Fig.~\ref{fig:ULX2PG}).

Fig.~\ref{fig:ULX2PG} shows that for obsID 10723 we can exclude an absorbed power law of $\Gamma\leq 2.2$ at $95\%$ confidence.  A source in the canonical hard state would be well described by a power law of $\Gamma \sim 1.7$, with $\Gamma=2.1$ at the most extreme for GRS 1915+105 \citep[based on values quoted in ][]{2006ARA&A..44...49R}.  However, some authors suggest that GRS 1915+105 does not experience the canonical hard state \citep{2003A&A...412..229R}, and the hardest spectra for the canonical hard state have $\Gamma \sim 1.9$, and tend to be for the sources that peak at lower luminosities.  therefore it seems unlikely that the source is in the BH LMXB low/hard state during this observation.  We find that $\Gamma=1.5$ is still allowed at the 20\%, 10\%, and 30\% levels for obsIDs 10722, 10725, and 12156 respectively, and $\sim10\%$ for ULX1 in obsID 10726, so for these detections, this method is inconclusive.   Therefore it seems that the source is not in the canonical hard state in at least one of the snapshot observations.  That the spectra appear to harden during the decline in luminosity is consistent with the transition from the thermal dominant to low/hard state, a fortiori because the hardest BH XB spectra are associated with this transition \cite{2008ApJ...682..212W}.

It is suggestive that the subsequent observation, obsID 10724, is the furthest median energy from obsID 10723, (table~\ref{tab:quan}) and has a median energy consistent with a grid of power law states (Fig.~\ref{fig:ULX2PG}).  Tentatively, this is indicative of the source moving from a soft, higher luminosity state, to a harder, lower luminosity state, as observed for the Galactic binaries as they return to quiescence.   Spectral fitting obsID 10722, for which we have the most counts from a single observation, showed no preference between power law or thermal dominant states.  However, fitting both spectral models recovered parameters consistent with those of the Galactic BH XBs.

\section{Discussion}
This work focuses on a series of regular \emph{Chandra} observations of Cen A to search for recurrence of the two known ULXs.  These sources are transient in nature, and only a handful of similar sources (observed $L{x} > \lerg{39}$ and $L_x < \lerg{38}$ in separate epochs) have been reported.   

The Cen A ULXs have been documented at $L_x >\lerg{39}$, with count rates of $(7-8) \times 10^{-2}~{ \rm s^{-1}}$ \citep{2006ApJ...640..459G,2008ApJ...677L..27S}.  By contrast, we present new detections in the $(0.5-5.0)\times 10^{-3}~{ \rm s^{-1}}$ range.  If one assumes a canonical hard state for all detections, i.e. a power law of $\Gamma\sim1.7$ experiencing Galactic absorption, this equates to luminosities $\approx10^{37}-\lerg{38}$.  For the low counts regime, estimating the exact luminosity will of course vary by a factor of a few depending on the assumed spectral model.  This is an opportunity to examine ULXs at substantially sub-Eddington luminosities, and compare them to Galactic BH XBs.

\subsection{Spectral States}
Galactic BH XBs typically emit the most energy as X-rays in the thermal state, where the contribution of comptonized emission to the spectrum is low. Physically, it is thought that the softer state results from the exposed optically thick, geometrically thin accretion disk, which 
is well-described by a disk blackbody.  In this model, the temperature $T_{in}$ of the innermost part of the accretion disk is recoverable by performing detailed spectral fitting, and for stellar mass BHs $kT_{in}$ $\sim 1.0$ keV \citep[but can reach $1.5$ keV, e.g.][]{2000ApJ...544..993S}.  The temperature of the disk scales inversely with radius from the center of the compact object, and as more mass leads to a larger ISCO, therefore more massive objects have cooler inner disks.  In terms of accretion rate $\dot{m}$, thermal states are observed between $0.03 -- 0.5$ Eddington for stellar mass BHs; however, it was noted by \cite{2009ApJ...695.1614S} that ULXs are not in the thermal state at peak luminosity, remaining hard during any variability.  A possible explanation is that such hard-state ULXs are IMBHs \citep{2006ApJ...649..730W}, that never reach accretion rates $\dot{m} \approx 0.03$, high enough to make them switch to the thermal state, even at their peak luminosities. Alternatively, if ULXs contain BHs of stellar origin ($M < 100 M_{\odot}$), their broad power law-like X-ray spectrum at peak luminosity may be the result of Comptonization in an optically thick medium \citep{2007Ap&SS.311..203R,2009MNRAS.397.1836G}; they would not be in the thermal dominant state because $\dot{m} \ga 1$.  In the latter scenario, we predict that the thermal state may be seen in transient sources during their decline, when $0.03 \la \dot{m} \la 0.5$, before they switch back to a hard state for $\dot{m} \la 0.03$. Failure to see transient ULXs passing through the thermal dominant state is either due to the very small number of such sources observed so far in the right accretion regime, or suggests that not all BHs necessarily switch to the thermal state during their outburst evolution.

The results from quantile analysis for the two transient ULXs in Cen A support the possibility that ULXs enter the canonical thermal state when they decline to lower luminosities.  Both sources have spectra consistent with those of XBs at $\lerg{38}$, and in one observation of ULX2, obsID 10723, we can exclude the hard state with $>95\%$ confidence based on the median energy of the spectrum, while analysis of the other spectra were inconclusive. At even lower luminosities, $L_x\sim 2\times \lerg{37}$, there is tentative evidence of a hardening of the ULX2 spectra, consistent with the behaviour of the canonical low/hard state.  No intra-observational variability was detected from these sources using the \emph{ciao} tool \emph{glvary}, as would be characteristic behaviour in the hard state; however, we are skeptical as to whether such variability would be detected, and note that it has previously not been detected using higher-quality data from hard state sources in Cen A \citep{2013ApJ...766...88B}.  These spectral results broadly support the argument that these ULXs contain stellar mass BHs.  

 The detection of ULX2 in a thermal state at $\sim\lerg{38}$ is consistent with other work on transient ULXs. A transient source in M31 that peaked at between $(2-5)\times\lerg{39}$ ULXs using \emph{XMM Newton}, had spectra that were described well by a disk blackbody model down to $L_x\approx 6\times\lerg{38}$ \citep[e.g.][]{2012MNRAS.420.2969M,2012A&A...538A..49K,2012MNRAS.tmp..186E}.  Steep power law spectra (interpreted as near-Eddington states) and soft states have been inferred for the globular cluster transient ULXs in NGC 4649 \citep[$\sim(2-5) \times \lerg{38}$,][]{2012ApJ...760..135R} and NGC 1399 \citep[$5 \times \lerg{38}$][]{2010ApJ...721..323S}, respectively. As a possible exception to this trend, a transient ULX in NGC 3379 \citep[][]{2008ApJS..179..142B} has been inferred progressing to a hard state at $\sim1\times10^{39}$ \citep[][]{2012arXiv1206.5304B}, and while the source was later detected at $\approx2\times\lerg{37}$, there were too few counts to determine if the spectra remained hard.

Based on our inferred spectral properties, we estimate the source luminosity based on the appropriate model.  Assuming that ULX2 is in a soft state in obsID 10723, we take the spectrum of the source to be dominated by a $kT_{in}=1.0$ keV disk blackbody experiencing Galactic absorption ($N_H=8.4\times {\rm 10^{20}~cm^{-2}~s^{-1} }$), then the observed $90\%$ confidence count rate corresponds to $L_x\sim (0.5-1.3)\times \lerg{38}$ in the $0.5-8.0$ keV  band.   In terms of systematic differences in the parameters of this model, changing  $kT_{in}$ to 0.5 keV only reduces the maximum luminosity by $\sim 10\%$.    ObsID 10723 has the highest count rate observed in the post-VLP observations, and therefore it is secure to state that our subsequent detections of both ULXs are below $L_x \sim \lerg{38}$.  Similarly we estimate the luminosity in obsID 10722 to be $\approx2 \times \lerg{37}$.

\subsection{Outburst Duration}
It is clear that both ULXs spend an appreciable amount of time in outburst.  ULX1 has now been detected in three separate epochs (1995 with \emph{ROSAT}, 1999 and 2011 with \emph{Chandra}) with secure non-detections between them  down to an interesting luminosity ($<\lerg{37}$).  In contrast, ULX2 remained undetected until 2007, and the number of outbursts that we have witnessed is less clear-cut.  There are two reasonable scenarios; the first is that we have witnessed a single outburst that was at UL luminosities in 2007, decayed from $\lerg{38}$ down to $\approx \lerg{37}$ between MJD 54835 and MJD 55082 (April-September 2009), and then, because this is approximately the detection limit for a 5 ks observation, remains on the cusp of being detected.  The second possibility is that there have been two outbursts, one in 2007 and another in 2009, of which we only witness the egress, and then the luminosity stays at $\approx\lerg{37}$ for hundreds of days. 

The overall behaviour of ULX2, while similar to other transient ULXs, is not directly analogous to any of the Galactic BH LMXBs.  In terms of the shape of the lightcurve, very few are seen to outburst longer than a few hundred days, and certainly not as long as a thousand \citep{2006ARA&A..44...49R}. However, some show signs of slow decays the end-points of which are unknown owing to the end of observations, such as V404 Cyg, which was probably still active after cessation of observations 150 days into its 1989 outburst \citep[e.g.,][]{1997ApJ...491..312C}.  The long period spent at lower luminosities for ULX2 is reminiscent of Swift J1753.5-0127 which has been seen to linger at lower luminosities an order of magnitude or more below the peak luminosity for $>1500$ days as of May 2010  \citep[RXTE, ][]{2012MNRAS.tmp..336S}, displaying both hard and soft spectral states during the lingering stage.  However, the peak luminosity of J1753.5-0127 is probably not much more than $\lerg{36}$, and therefore the low/hard state is close to what would be considered quiescence for a lot of sources.  The comparably low luminosity may be a consequence of the tightness of this particular system ($P_{orb}\sim 3.2$~hr).   

Alternatively, it is perhaps only the persistent systems with higher mass companions that would potentially give similar lightcurves if sampled with \emph{Chandra} from Cen A, such as Cyg X-1, LMC X-1 and LMC X-3.  The transient source GRS 1915+105, which has been in outburst since 1992, could be a more pertinent comparison, depending on whether ULX2 has now returned to quiescence or is still in outburst, below $\lerg{37}$.

\subsection{Duty Cycle}
\label{sec:cycle}
The duty cycle $d$ is defined as the ratio of the time spent in outburst of the total time spent in outburst and quiessence.

For ULX1, let us assume that an outburst lasts a maximum of $\approx 500$ days (Fig.~\ref{fig:LC}), which is consistent with the $\approx 800$ day upper limit of \cite{2000A&A...357L..57S}, and also assume that we have observed every possible outburst (3) since 1990.    Then it would seem that a reasonable upper limit for $d_{ULX1}\approx1500/8000\sim0.19$, as the outbursts will probably be shorter than 500 days, which compensates for the possibility of there having been unobserved outbursts. 

ULX2 was first detected at the start of the VLP (MJD 54181) and the outburst lasts at least until the end of obsID 12156 (MJD 55552), assuming a single outburst. This places a secure upper limit on the time spent in outburst of at least 1553 days (Fig.~\ref{fig:LC}). Under the two-outburst scenario, the shortest possible outburst durations are 69 days (MJD 54228 from MJD 54181) plus 899 days (MJD 55734 from MJD 54835) in the second outburst, giving 968 days in outburst over the 20 years of observations. For ULX2, the longest reasonable $\tau_{out}\approx1600$ days, yielding $d_{ULX2}\sim 0.2$, similar to ULX1, while a lower limit based on $\sim 970$ days observed in outburst gives $d_{ULX2}\approx0.12$.

To emit persistently at UL luminosities, a low-mass donor would be exhausted on timescales $~10^{6}$ years after the initial mass transfer is triggered, assuming radiative efficiency $\eta=0.1$.  However, the Cen A sources reach the ULX threshold only for a fraction of the time, their luminosity averaged over the the outburst phases is only $\sim \lerg{37}$, corresponding to $\dot{M}\sim 10^{-9} {~M_{\odot}~ \rm yr^{-1}}$.  If they do have duty cycles as high as $20\%$, this implies that their donor stars can sustain them for a characteristic timescale of $~1$ Gyr after the onset of mass transfer.  Based on this characteristic behaviour, we speculate that ULX behaviour is associated with an evolved donor star (subgiant and red giant phases).

\subsection{Other Early-type Galaxies}
\label{sec:othergal}
We investigate an apparent paucity of ULX transient candidates in other early-type galaxies compared to Cen A.  Only two other  such sources are known, which are globular cluster ULXs in NGC 3379 and NGC 4649 \citep{2012arXiv1206.5304B,2012ApJ...760..135R}. \cite{2012A&A...546A..36Z} showed there are approximately $12-13$ ULXs in a sample of 20 large early-type galaxies that included NGC 4649 and NGC 3379 but excluded Cen A, suggesting that transient candidates typically make up only $\sim 15\%$ of the ULXs in early-type galaxies.  To evaluate the unusualness of the lack of discoveries we examine the large ellipticals NGC 4472 and NGC 4649, which have 171 and 168 XBs above a limiting luminosity of $L_x\sim6\times\lerg{37}$ \citep[completeness corrected, see ][]{2012A&A...546A..36Z}.  By contrast, \cite{2013ApJ...766...88B} showed that Cen A has $\sim 35$ sources above this luminosity, two of which are ULX transients\footnote[5]{\cite{2013ApJ...766...88B} studied sources inside the half-light radius of Cen A, whereas Zhang et al. looked at sources within the $D_{25}$ ellipses; however, \cite{2006A&A...447...71V} showed that the background sources became the major contributor to the point source population of Cen A at the half-light radius, with only $\approx10$ LMXBs beyond this point. \cite{2012A&A...546A..36Z} corrected for background sources in their population study, so it is valid to compare these numbers.}.    Scaling the number of ULX per number of LMXBs leads to an expectation of $\approx10$ transient ULXs per galaxy.  To date, only one confirmed ULX with transient-like behaviour has been observed in either \citep[in NGC 4649, ][]{2012ApJ...760..135R}.

\begin{figure}
\begin{center}
\includegraphics[height=0.48\textwidth, angle=270]{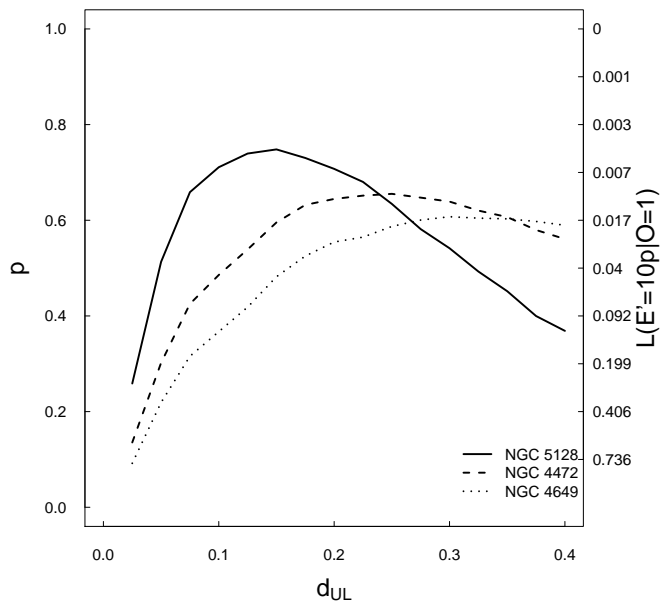}
\caption{Proportion of simulated transient ULX detections for different UL duty cycles given the \emph{Chandra} observing campaign for galaxies NGC 4472 and NGC 4649 and NGC 5128 (Cen A).  Secondary y-axis shows probability of observing one or fewer such sources, when there are ten per galaxy, using expectation value $10 p$ (see \S \ref{sec:othergal}). \label{fig:probs}}
\end{center}
\end{figure}

Is it strange that only one such source has been detected given that we expect each galaxy to possess $\sim10$? To answer this we have to consider the respective observing campaigns for NGC 4472 and NGC 4649.  We simulated $10^{5}$ on-off lightcurves for a range of duty cycles, where we define an `ultraluminous' duty cycle $d_{UL}$, which only considers the period of time spent $L_x > \lerg{39}$ (i.e. where `on' corresponds to $L_x>\lerg{39}$ and `off' represents some less luminous epoch) and folded these through the respective \emph{Chandra} observing campaigns.  To register as a transient ULX, a source had to be observed in both regimes by the sequence of observations.  In Fig.~\ref{fig:probs} we present the proportion $p$ of simulated lightcurves that registered as transient ULXs against $d_{UL}$.  To judge how unusual a detection of one source in a given galaxy is, we take an expectation value $E_{ulx}(=10)$ modified by $p$ such that $E^{\prime}_{ulx} = p E_{ulx}$ (e.g. in Fig.~\ref{fig:probs}, for NGC 4649 where $E_{ulx}=10$, $E^{\prime}< 6$ for all $d_{UL}$).  For an observed number of transient ULXs $O_{ulx}$ given an expectation number $E^{\prime}_{ulx}$ we assessed the likelihood function $\mathcal{L}(E^{\prime}_{ulx} | O_{ulx}) \equiv P(O_{ulx} | E^{\prime}_{ulx})$ for $O_{ulx}=1$.  This is simply the poisson probability where $\lambda = E^{\prime}_{ulx}$, which we plot as a secondary y-axis in Fig.~\ref{fig:probs}, taking $E_{ulx}=10$.

For NGC 4472, Fig.~\ref{fig:probs} shows that a detection of 1 or fewer
sources is indeed unusual at $95\%$ confidence for $d_{UL}\gtrsim0.1$, and
for NGC 4649, this is closer to 0.2.  These values are consistent with the
outburst duty cycle $d$ discussed in \S\ref{sec:cycle}, and by definition
$d_{ul}\le d$.  Therefore, at least two possibilities exist to account for 
the scarcity of similar sources identified in NGC 4472 and NGC 4649.
Firstly, the duty cycle of the UL regime may be small, $\lesssim10\%$,
which would be consistent with our results, and detections unlikely from
the current set of observations.  Secondly, it may be wrong to assume 
that the number of transient ULXs scales with the total number of LMXBs. There may be other 
properties of the stellar populations that are enhancing the creation 
of transient ULXs in Cen A compared with those other two galaxies. 
In particular, the evolutionary stage of the donor star may be a key 
factor. It has been tentatively suggested that ULX2 has a K giant 
companion~\citep{2008ApJ...677L..27S}; more generally, we speculate 
that transient ULXs in an old population are mostly associated 
with low-mass donors in the subgiant or giant stage, because those 
evolved donors are more likely to provide the mass transfer rate 
required to achieve ULX luminosity. If so, the number of ULX 
transients in a galaxy may be proportional to the number of evolved 
stars, which is a function of both age and total stellar mass.
The stellar mass of Cen A ($M_{\ast} \approx (1-2)\times 10^{11} 
M_{\odot}$) is almost three times lower than the stellar masses 
of NGC 4472 and NGC 4649 ($M_{\ast} \approx 3.2 \times 10^{11} 
M_{\odot}$), based on their K-band brightnesses 
\citep{2006MNRAS.367..627E} and stellar mass-to-light ratios 
\citep{2006ApJ...646..899H}, and assuming a Kroupa IMF 
\citep{2001MNRAS.322..231K}. However, $\approx 20-30\%$ of the stars 
in the Cen A halo are $\approx4$ Gyr old \citep{2011A&A...526A.123R} 
by virtue of its more recent merger, while the characteristic age 
of the stellar population in the other two galaxies is $\sim 10$ Gyr.
Using Starburst99 \citep{1999ApJS..123....3L,2005ApJ...621..695V,2010ApJS..189..309L}, we estimate that NGC 4472 and NGC 4649 
currently contain $\approx 1.2 \times 10^{9}$ subgiants and giants each, 
while Cen A contains $\approx (0.6-1.2) \times10^{9}$ evolved stars.
Therefore, in this scenario $E_{ulx}$ is similar in all three galaxies 
and a non-discovery is consistent with the small-number statistics.

Regular, short observations of all three galaxies over an extended period of time would allow for a better determination of $d_{UL}$ and distinction between the effects of observational bias and host galaxy in the detection of transient ULX sources.

\section{Conclusion}
We report on the latest detections of the two known ULXs in Cen A and present the count rate lightcurves in Fig.\ref{fig:LC}.  These sources are at substantially sub-Eddington luminosities.  We place the duty cycles in the range $12-20\%$ and it seems likely that much of the time in outburst is spent at luminosities of a few $\lerg{37}$.  Both sources are currently at luminosities below $\lerg{37}$, based on combined limits from the three most recent observations, and may have gone into a quiescent state.

The available evidence from studying the spectra of these sources at lower luminosity favours stellar mass BH primaries that occasionally emit at Eddington/super-Eddington isotropic luminosities, rather than IMBH accretors.  This is consistent with other recent studies of transient ULXs at sub-Eddington luminosities.  

We attempt to account for an apparent lack of transient ULXs in other, much larger early-type galaxies.  While this could be accounted for by the majority of such sources spending only a small amount of their outbursts in the UL regime, our total knowledge of transient ULXs suggests that it is not uncommon for them to spend over a year at such luminosities.  It seems likely that the number of transient ULXs scales not with the XB population, but with the size of the subgiant and giant populations, which are comparable between Cen A and the larger galaxies.  This is consistent with the tentative suggestion that ULX2 has a K giant companion.

This work was supported by NASA grant NAS8-03060. MJB thanks SAO and the University of Birmingham for financial support. GRS acknowledges the support of an NSERC Discovery Grant. We also thank Jeanette Gladstone, Chris Done and Ewan O'Sullivan for useful discussions. We extend our thanks to the anonymous referee for useful comments that improved the paper.
\bibliography{ulxbib}{}

\begin{thebibliography}{65}
\expandafter\ifx\csname natexlab\endcsname\relax\def\natexlab#1{#1}\fi

\bibitem[{{Arnaud} {et~al.}(2011){Arnaud}, {Smith}, \&
  {Siemiginowska}}]{2011hxa..book.....A}
{Arnaud}, K., {Smith}, R., \& {Siemiginowska}, A. 2011, {Handbook of X-ray
  Astronomy}, ed. R.~{Ellis}, J.~{Huchra}, S.~{Kahn}, G.~{Rieke}, \& P.~B.
  {Stetson}

\bibitem[{{Bauer} \& {Pietsch}(2005)}]{2005A&A...442..925B}
{Bauer}, M., \& {Pietsch}, W. 2005, \aap, 442, 925

\bibitem[{{Brassington} {et~al.}(2008){Brassington}, {Fabbiano}, {Kim},
  {Zezas}, {Zepf}, {Kundu}, {Angelini}, {Davies}, {Gallagher}, {Kalogera},
  {Fragos}, {King}, {Pellegrini}, \& {Trinchieri}}]{2008ApJS..179..142B}
{Brassington}, N.~J., {Fabbiano}, G., {Kim}, D.-W., {et~al.} 2008, \apjs, 179,
  142

\bibitem[{{Brassington} {et~al.}(2010){Brassington}, {Fabbiano}, {Blake},
  {Zezas}, {Angelini}, {Davies}, {Gallagher}, {Kalogera}, {Kim}, {King},
  {Kundu}, {Trinchieri}, \& {Zepf}}]{2010ApJ...725.1805B}
{Brassington}, N.~J., {Fabbiano}, G., {Blake}, S., {et~al.} 2010, \apj, 725,
  1805

\bibitem[{{Brassington} {et~al.}(2012){Brassington}, {Fabbiano}, {Zezas},
  {Kundu}, {Kim}, {Fragos}, {King}, {Pellegrini}, {Trinchieri}, {Zepf}, \&
  {Wright}}]{2012arXiv1206.5304B}
{Brassington}, N.~J., {Fabbiano}, G., {Zezas}, A., {et~al.} 2012, \apj, 755,
  162

\bibitem[{{Burke} {et~al.}(2012){Burke}, {Raychaudhury}, {Kraft},
  {Brassington}, {Hardcastle}, {Goodger}, {Sivakoff}, {Forman}, {Jones},
  {Woodley}, {Murray}, {Kainulainen}, {Birkinshaw}, {Croston}, {Evans},
  {Gilfanov}, {Jord{\'a}n}, {Sarazin}, {Voss}, {Worrall}, \&
  {Zhang}}]{2012ApJ...749..112B}
{Burke}, M.~J., {Raychaudhury}, S., {Kraft}, R.~P., {et~al.} 2012, \apj, 749,
  112

\bibitem[{{Burke} {et~al.}(2013){Burke}, {Raychaudhury}, {Kraft}, {Maccarone},
  {Brassington}, {Hardcastle}, {Kainulainen}, {Woodley}, {Goodger}, {Sivakoff},
  {Forman}, {Jones}, {Murray}, {Birkinshaw}, {Croston}, {Evans}, {Gilfanov},
  {Jord{\'a}n}, {Sarazin}, {Voss}, {Worrall}, \& {Zhang}}]{2013ApJ...766...88B}
---. 2013, \apj, 766, 88

\bibitem[{{Cash}(1979)}]{1979ApJ...228..939C}
{Cash}, W. 1979, \apj, 228, 939

\bibitem[{{Chen} {et~al.}(1997){Chen}, {Shrader}, \&
  {Livio}}]{1997ApJ...491..312C}
{Chen}, W., {Shrader}, C.~R., \& {Livio}, M. 1997, \apj, 491, 312

\bibitem[{{Colbert}(2004)}]{2004AAS...204.4904C}
{Colbert}, E.~J.~M. 2004, in Bulletin of the American Astronomical Society,
  Vol.~36, American Astronomical Society Meeting Abstracts \#204, 749

\bibitem[{{Constantin} {et~al.}(2009){Constantin}, {Green}, {Aldcroft}, {Kim},
  {Haggard}, {Barkhouse}, \& {Anderson}}]{2009ApJ...705.1336C}
{Constantin}, A., {Green}, P., {Aldcroft}, T., {et~al.} 2009, \apj, 705, 1336

\bibitem[{{Coriat} {et~al.}(2012){Coriat}, {Fender}, \&
  {Dubus}}]{2012MNRAS.424.1991C}
{Coriat}, M., {Fender}, R.~P., \& {Dubus}, G. 2012, \mnras, 424, 1991

\bibitem[{{Dubus} {et~al.}(2001){Dubus}, {Hameury}, \&
  {Lasota}}]{2001A&A...373..251D}
{Dubus}, G., {Hameury}, J.-M., \& {Lasota}, J.-P. 2001, \aap, 373, 251

\bibitem[{{Ellis} \& {O'Sullivan}(2006)}]{2006MNRAS.367..627E}
{Ellis}, S.~C., \& {O'Sullivan}, E. 2006, \mnras, 367, 627

\bibitem[{{Esposito} {et~al.}(2012){Esposito}, {Motta}, {Pintore}, {Zampieri},
  \& {Tomasella}}]{2012MNRAS.tmp..186E}
{Esposito}, P., {Motta}, S.~E., {Pintore}, F., {Zampieri}, L., \& {Tomasella},
  L. 2012, \mnras, 186

\bibitem[{Feng \& Kaaret(2007)}]{0004-637X-668-2-941}
Feng, H., \& Kaaret, P. 2007, The Astrophysical Journal, 668, 941

\bibitem[{{Ferrarese} {et~al.}(2007){Ferrarese}, {Mould}, {Stetson}, {Tonry},
  {Blakeslee}, \& {Ajhar}}]{2007ApJ...654..186F}
{Ferrarese}, L., {Mould}, J.~R., {Stetson}, P.~B., {et~al.} 2007, \apj, 654,
  186

\bibitem[{{Ghosh} {et~al.}(2006){Ghosh}, {Finger}, {Swartz}, {Tennant}, \&
  {Wu}}]{2006ApJ...640..459G}
{Ghosh}, K.~K., {Finger}, M.~H., {Swartz}, D.~A., {Tennant}, A.~F., \& {Wu}, K.
  2006, \apj, 640, 459

\bibitem[{{Gladstone} {et~al.}(2013){Gladstone}, {Copperwheat}, {Heinke},
  {Roberts}, {Cartwright}, {Levan}, {Goad}, \& {.}}]{2013arXiv1303.1213G}
{Gladstone}, J.~C., {Copperwheat}, C., {Heinke}, C.~O., {et~al.} 2013, ArXiv
  e-prints

\bibitem[{{Gladstone} {et~al.}(2009){Gladstone}, {Roberts}, \&
  {Done}}]{2009MNRAS.397.1836G}
{Gladstone}, J.~C., {Roberts}, T.~P., \& {Done}, C. 2009, \mnras, 397, 1836

\bibitem[{{Gregory} \& {Loredo}(1992)}]{1992ApJ...398..146G}
{Gregory}, P.~C., \& {Loredo}, T.~J. 1992, \apj, 398, 146

\bibitem[{{Hong} {et~al.}(2004){Hong}, {Schlegel}, \&
  {Grindlay}}]{2004ApJ...614..508H}
{Hong}, J., {Schlegel}, E.~M., \& {Grindlay}, J.~E. 2004, \apj, 614, 508

\bibitem[{{Hong} {et~al.}(2009){Hong}, {van den Berg}, {Grindlay}, \&
  {Laycock}}]{2009ApJ...706..223H}
{Hong}, J.~S., {van den Berg}, M., {Grindlay}, J.~E., \& {Laycock}, S. 2009,
  \apj, 706, 223

\bibitem[{{Humphrey} {et~al.}(2006){Humphrey}, {Buote}, {Gastaldello},
  {Zappacosta}, {Bullock}, {Brighenti}, \& {Mathews}}]{2006ApJ...646..899H}
{Humphrey}, P.~J., {Buote}, D.~A., {Gastaldello}, F., {et~al.} 2006, \apj, 646,
  899

\bibitem[{{Irwin}(2006)}]{2006MNRAS.371.1903I}
{Irwin}, J.~A. 2006, \mnras, 371, 1903

\bibitem[{{Jord{\'a}n} {et~al.}(2007){Jord{\'a}n}, {Sivakoff}, {McLaughlin},
  {Blakeslee}, {Evans}, {Kraft}, {Hardcastle}, {Peng}, {C{\^o}t{\'e}},
  {Croston}, {Juett}, {Minniti}, {Raychaudhury}, {Sarazin}, {Worrall},
  {Harris}, {Woodley}, {Birkinshaw}, {Brassington}, {Forman}, {Jones}, \&
  {Murray}}]{2007ApJ...671L.117J}
{Jord{\'a}n}, A., {Sivakoff}, G.~R., {McLaughlin}, D.~E., {et~al.} 2007, \apjl,
  671, L117

\bibitem[{{Kalogera} {et~al.}(2004){Kalogera}, {Henninger}, {Ivanova}, \&
  {King}}]{2004ApJ...603L..41K}
{Kalogera}, V., {Henninger}, M., {Ivanova}, N., \& {King}, A.~R. 2004, \apjl,
  603, L41

\bibitem[{{Kaur} {et~al.}(2012){Kaur}, {Henze}, {Haberl}, {Pietsch}, {Greiner},
  {Rau}, {Hartmann}, {Sala}, \& {Hernanz}}]{2012A&A...538A..49K}
{Kaur}, A., {Henze}, M., {Haberl}, F., {et~al.} 2012, \aap, 538, A49

\bibitem[{{King} {et~al.}(2001){King}, {Davies}, {Ward}, {Fabbiano}, \&
  {Elvis}}]{2001ApJ...552L.109K}
{King}, A.~R., {Davies}, M.~B., {Ward}, M.~J., {Fabbiano}, G., \& {Elvis}, M.
  2001, \apjl, 552, L109

\bibitem[{{Kraft} {et~al.}(2001){Kraft}, {Kregenow}, {Forman}, {Jones}, \&
  {Murray}}]{2001ApJ...560..675K}
{Kraft}, R.~P., {Kregenow}, J.~M., {Forman}, W.~R., {Jones}, C., \& {Murray},
  S.~S. 2001, \apj, 560, 675

\bibitem[{{Kreidberg} {et~al.}(2012){Kreidberg}, {Bailyn}, {Farr}, \&
  {Kalogera}}]{2012ApJ...757...36K}
{Kreidberg}, L., {Bailyn}, C.~D., {Farr}, W.~M., \& {Kalogera}, V. 2012, \apj,
  757, 36

\bibitem[{{Kroupa}(2001)}]{2001MNRAS.322..231K}
{Kroupa}, P. 2001, \mnras, 322, 231

\bibitem[{{Lasota}(2001)}]{2001NewAR..45..449L}
{Lasota}, J.-P. 2001, NAR, 45, 449

\bibitem[{{Leitherer} {et~al.}(2010){Leitherer}, {Ortiz Ot{\'a}lvaro},
  {Bresolin}, {Kudritzki}, {Lo Faro}, {Pauldrach}, {Pettini}, \&
  {Rix}}]{2010ApJS..189..309L}
{Leitherer}, C., {Ortiz Ot{\'a}lvaro}, P.~A., {Bresolin}, F., {et~al.} 2010,
  \apjs, 189, 309

\bibitem[{{Leitherer} {et~al.}(1999){Leitherer}, {Schaerer}, {Goldader},
  {Gonz{\'a}lez Delgado}, {Robert}, {Kune}, {de Mello}, {Devost}, \&
  {Heckman}}]{1999ApJS..123....3L}
{Leitherer}, C., {Schaerer}, D., {Goldader}, J.~D., {et~al.} 1999, \apjs, 123,
  3

\bibitem[{{Liu} {et~al.}(2006){Liu}, {Bregman}, \&
  {Irwin}}]{2006ApJ...642..171L}
{Liu}, J.-F., {Bregman}, J.~N., \& {Irwin}, J. 2006, \apj, 642, 171

\bibitem[{{Liu} {et~al.}(2004){Liu}, {Bregman}, \&
  {Seitzer}}]{2004ApJ...602..249L}
{Liu}, J.-F., {Bregman}, J.~N., \& {Seitzer}, P. 2004, \apj, 602, 249

\bibitem[{{Maccarone}(2003)}]{2003A&A...409..697M}
{Maccarone}, T.~J. 2003, \aap, 409, 697

\bibitem[{{Maccarone} {et~al.}(2003){Maccarone}, {Gallo}, \&
  {Fender}}]{2003MNRAS.345L..19M}
{Maccarone}, T.~J., {Gallo}, E., \& {Fender}, R. 2003, \mnras, 345, L19

\bibitem[{{Middleton} {et~al.}(2012){Middleton}, {Sutton}, {Roberts},
  {Jackson}, \& {Done}}]{2012MNRAS.420.2969M}
{Middleton}, M.~J., {Sutton}, A.~D., {Roberts}, T.~P., {Jackson}, F.~E., \&
  {Done}, C. 2012, \mnras, 420, 2969

\bibitem[{{{\"O}zel} {et~al.}(2010){{\"O}zel}, {Psaltis}, {Narayan}, \&
  {McClintock}}]{2010ApJ...725.1918O}
{{\"O}zel}, F., {Psaltis}, D., {Narayan}, R., \& {McClintock}, J.~E. 2010,
  \apj, 725, 1918

\bibitem[{{Park} {et~al.}(2006){Park}, {Kashyap}, {Siemiginowska}, {van Dyk},
  {Zezas}, {Heinke}, \& {Wargelin}}]{2006ApJ...652..610P}
{Park}, T., {Kashyap}, V.~L., {Siemiginowska}, A., {et~al.} 2006, \apj, 652,
  610

\bibitem[{{Piro} \& {Bildsten}(2002)}]{2002ApJ...571L.103P}
{Piro}, A.~L., \& {Bildsten}, L. 2002, \apjl, 571, L103

\bibitem[{{Portegies Zwart} {et~al.}(2004){Portegies Zwart}, {Dewi}, \&
  {Maccarone}}]{2004MNRAS.355..413P}
{Portegies Zwart}, S.~F., {Dewi}, J., \& {Maccarone}, T. 2004, \mnras, 355, 413

\bibitem[{{Reig} {et~al.}(2003){Reig}, {Belloni}, \& {van der
  Klis}}]{2003A&A...412..229R}
{Reig}, P., {Belloni}, T., \& {van der Klis}, M. 2003, \aap, 412, 229

\bibitem[{{Rejkuba} {et~al.}(2011){Rejkuba}, {Harris}, {Greggio}, \&
  {Harris}}]{2011A&A...526A.123R}
{Rejkuba}, M., {Harris}, W.~E., {Greggio}, L., \& {Harris}, G.~L.~H. 2011,
  \aap, 526, A123

\bibitem[{{Remillard} \& {McClintock}(2006)}]{2006ARA&A..44...49R}
{Remillard}, R.~A., \& {McClintock}, J.~E. 2006, \araa, 44, 49

\bibitem[{{Roberts}(2007)}]{2007Ap&SS.311..203R}
{Roberts}, T.~P. 2007, \apss, 311, 203

\bibitem[{{Roberts} {et~al.}(2012){Roberts}, {Fabbiano}, {Luo}, {Kim},
  {Strader}, {Middleton}, {Brodie}, {Fragos}, {Gallagher}, {Kalogera}, {King},
  \& {Zezas}}]{2012ApJ...760..135R}
{Roberts}, T.~P., {Fabbiano}, G., {Luo}, B., {et~al.} 2012, \apj, 760, 135

\bibitem[{{Servillat} {et~al.}(2011){Servillat}, {Farrell}, {Lin}, {Godet},
  {Barret}, \& {Webb}}]{2011ApJ...743....6S}
{Servillat}, M., {Farrell}, S.~A., {Lin}, D., {et~al.} 2011, \apj, 743, 6

\bibitem[{{Shih} {et~al.}(2010){Shih}, {Kundu}, {Maccarone}, {Zepf}, \&
  {Joseph}}]{2010ApJ...721..323S}
{Shih}, I.~C., {Kundu}, A., {Maccarone}, T.~J., {Zepf}, S.~E., \& {Joseph},
  T.~D. 2010, \apj, 721, 323

\bibitem[{{Sivakoff} {et~al.}(2008){Sivakoff}, {Kraft}, {Jord{\'a}n}, {Juett},
  {Evans}, {Forman}, {Hardcastle}, {Sarazin}, {Birkinshaw}, {Brassington},
  {Croston}, {Harris}, {Jones}, {Murray}, {Raychaudhury}, {Woodley}, \&
  {Worrall}}]{2008ApJ...677L..27S}
{Sivakoff}, G.~R., {Kraft}, R.~P., {Jord{\'a}n}, A., {et~al.} 2008, \apjl, 677,
  L27

\bibitem[{{Smak}(1984)}]{1984PASP...96....5S}
{Smak}, J. 1984, \pasp, 96, 5

\bibitem[{{Sobczak} {et~al.}(2000){Sobczak}, {McClintock}, {Remillard}, {Cui},
  {Levine}, {Morgan}, {Orosz}, \& {Bailyn}}]{2000ApJ...544..993S}
{Sobczak}, G.~J., {McClintock}, J.~E., {Remillard}, R.~A., {et~al.} 2000, \apj,
  544, 993

\bibitem[{{Soleri} {et~al.}(2012){Soleri}, {Mu{\~n}oz-Darias}, {Motta},
  {Belloni}, {Casella}, {M{\'e}ndez}, {Altamirano}, {Linares}, {Wijnands},
  {Fender}, \& {van der Klis}}]{2012MNRAS.tmp..336S}
{Soleri}, P., {Mu{\~n}oz-Darias}, T., {Motta}, S., {et~al.} 2012, \mnras, 336

\bibitem[{{Soria} {et~al.}(2012){Soria}, {Kuntz}, {Winkler}, {Blair}, {Long},
  {Plucinsky}, \& {Whitmore}}]{2012ApJ...750..152S}
{Soria}, R., {Kuntz}, K.~D., {Winkler}, P.~F., {et~al.} 2012, \apj, 750, 152

\bibitem[{{Soria} {et~al.}(2009){Soria}, {Risaliti}, {Elvis}, {Fabbiano},
  {Bianchi}, \& {Kuncic}}]{2009ApJ...695.1614S}
{Soria}, R., {Risaliti}, G., {Elvis}, M., {et~al.} 2009, \apj, 695, 1614

\bibitem[{{Steinle} {et~al.}(2000){Steinle}, {Dennerl}, \&
  {Englhauser}}]{2000A&A...357L..57S}
{Steinle}, H., {Dennerl}, K., \& {Englhauser}, J. 2000, \aap, 357, L57

\bibitem[{{Swartz} {et~al.}(2009){Swartz}, {Tennant}, \&
  {Soria}}]{2009ApJ...703..159S}
{Swartz}, D.~A., {Tennant}, A.~F., \& {Soria}, R. 2009, \apj, 703, 159

\bibitem[{{V{\'a}zquez} \& {Leitherer}(2005)}]{2005ApJ...621..695V}
{V{\'a}zquez}, G.~A., \& {Leitherer}, C. 2005, \apj, 621, 695

\bibitem[{{Voss} \& {Gilfanov}(2006)}]{2006A&A...447...71V}
{Voss}, R., \& {Gilfanov}, M. 2006, \aap, 447, 71

\bibitem[{{Winter} {et~al.}(2006){Winter}, {Mushotzky}, \&
  {Reynolds}}]{2006ApJ...649..730W}
{Winter}, L.~M., {Mushotzky}, R.~F., \& {Reynolds}, C.~S. 2006, \apj, 649, 730

\bibitem[{{Wu} \& {Gu}(2008)}]{2008ApJ...682..212W}
{Wu}, Q., \& {Gu}, M. 2008, \apj, 682, 212

\bibitem[{{Wu} {et~al.}(2010){Wu}, {Yu}, {Li}, {Maccarone}, \&
  {Li}}]{2010ApJ...718..620W}
{Wu}, Y.~X., {Yu}, W., {Li}, T.~P., {Maccarone}, T.~J., \& {Li}, X.~D. 2010,
  \apj, 718, 620

\bibitem[{{Zhang} {et~al.}(2012){Zhang}, {Gilfanov}, \&
  {Bogd{\'a}n}}]{2012A&A...546A..36Z}
{Zhang}, Z., {Gilfanov}, M., \& {Bogd{\'a}n}, {\'A}. 2012, \aap, 546, A36

\end{thebibliography}
\bibliographystyle{apj}


\end{document}